\title{A Two Step Approach to Weighted Bipartite Link Recommendations}

\author{Nathan Ma}
\date{Adlai E. Stevenson High School, Lincolnshire, IL}
\newcommand{\abstractText}{\noindent
Many real world person-person or person-product relationships can be modeled graphically. More specifically, bipartite graphs can be especially useful when modeling scenarios that involve two disjoint groups. As a result, many existing papers have utilized bipartite graphs for the classical link recommendation problem. In this paper, using the principle of bipartite graphs, we present another approach to this problem with a two step algorithm that takes into account frequency and similarity between common edges to make recommendations. We test this approach with bipartite data gathered from the Epinions and Movielens data sources, and find it to perform with roughly 14 percent error, which improves upon baseline results. This is a promising result, and can be refined to generate even more accurate recommendations.
}
\newcommand{\sentence}{\noindent
This was coded in the PyCharm virtual environment, using the NetWorkX \cite{networkx} package. The below code is specific for how we loaded in and ran out algorithm on the Movielens dataset \cite{movielens}.
}

\documentclass[12pt]{article}
\usepackage{graphicx}
\usepackage{xurl}
\usepackage{indentfirst}
\usepackage{setspace}
\usepackage{amsmath}
\usepackage{float}
\usepackage{biblatex}
\usepackage{abstract}

\usepackage{lipsum}
\usepackage{listings}
\usepackage{geometry}
\usepackage{hyperref}
\hypersetup{colorlinks=true, urlcolor=blue, linkcolor=blue, citecolor=blue}

\begin{document}


    \maketitle
    
    \begin{abstract}
      \abstractText
      \newline
      \newline
    \end{abstract}


\section{Introduction}

Bipartite graphs \cite{Read_Atlas} are graphs that can be split into two sets of vertices such that, within each set, there are no edges. Edges only exist between vertices of opposite set. See Figure 1 for an example of a bipartite graph. Within a weighted bipartite graph, each edge is also given a weight.

Weighted bipartite graphs \cite{graph} can be used to model many real world relationships. For example, we can use a weighted bipartite graph to model the product-customer relation. In this situation, the weighted edges represent how a customer has rated a product. 

The link prediction problem \cite{social} takes in an arbitrary graph network and predicts the possibility of an edge between unpaired vertices. The link prediction problem isn't solely limited to weighted bipartite graphs. However, this paper focuses on examining the link prediction problem in the context of weighted bipartite graphs. In the case of the product-customer relationship, the link prediction problem is analogous to providing recommendations to customers on future purchases.

We answer the question of weighted link prediction with a two-step approach. First, we evaluate our confidence in a potential link occurring between two vertices based on their common neighbors. Second, we determine their weight based on the information these common neighbors provide. We test this algorithm on existing datasets that can be modeled with a weighted bipartite graph.

\begin{figure}
    \centering
    \includegraphics[width=6cm]{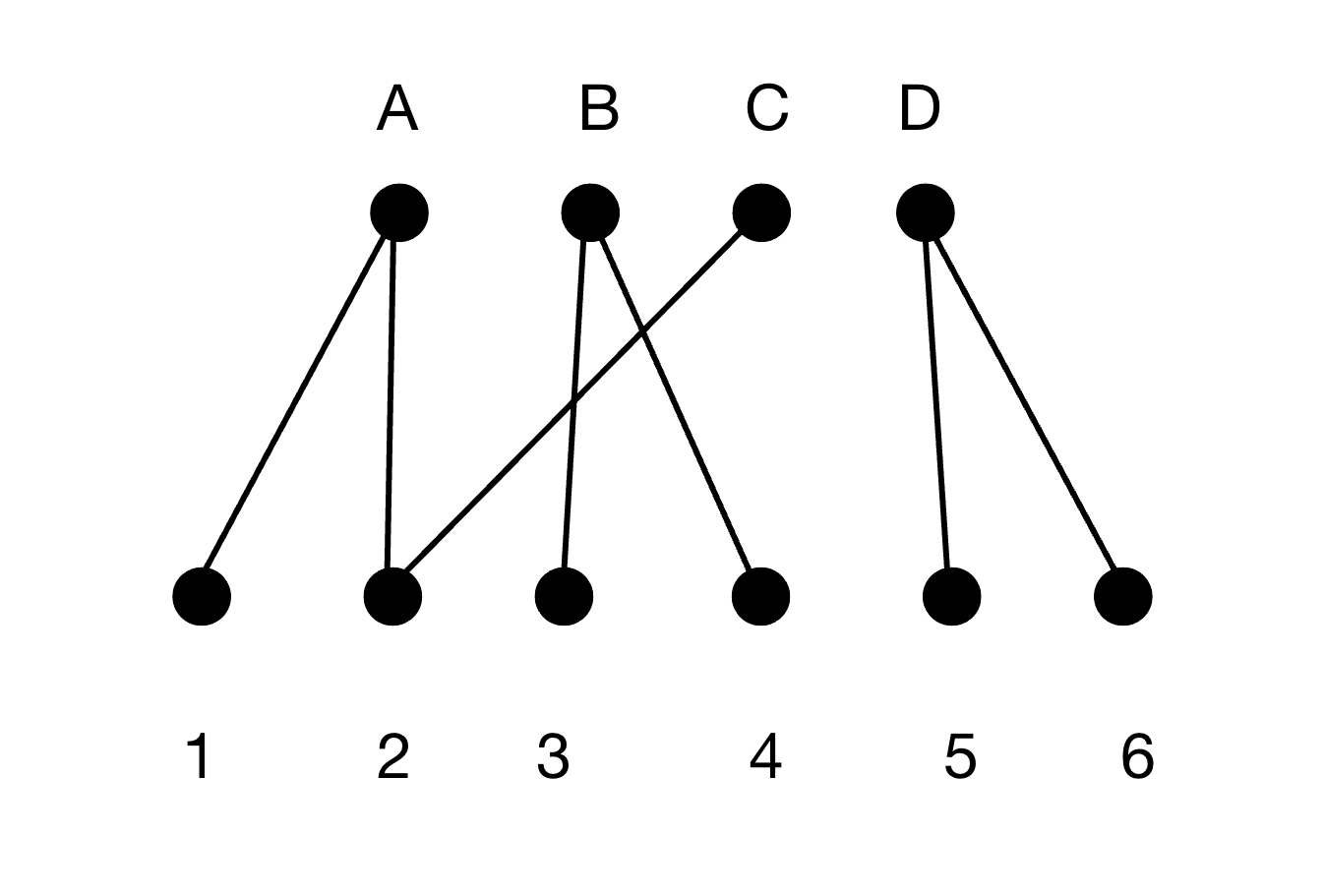}
    \caption{Example of an unweighted bipartite graph.}
\end{figure}

\section{Related Work}

Since link prediction has many real world applications, several papers \cite{LinkReference1}\cite{neural}\cite{jester} have already approached it. Some of these papers dealt with unweighted bipartite graphs \cite{LinkReference1} and others dealt with weighted bipartite graphs \cite{jester}.

One approach for unweighted link prediction has been to specifically predict internal links \cite{LinkReference1}. Internal links are defined as such: an unpaired top and bottom node such that a hypothetical edge between them would not change the bipartite projected graph. Only internal links were considered as potential edges because being "internally linked" means that two vertices already share some degree of common neighbors. This means that the two vertices are more likely to have an edge develop between them. Then, whether an internal link would become an actual edge is determined from several proposed weighting functions that took into account factors such as common neighbors and overall number of neighbors. See Figure 2 for an example of an internal link.

\begin{figure}
    \centering
    \includegraphics[width=6cm]{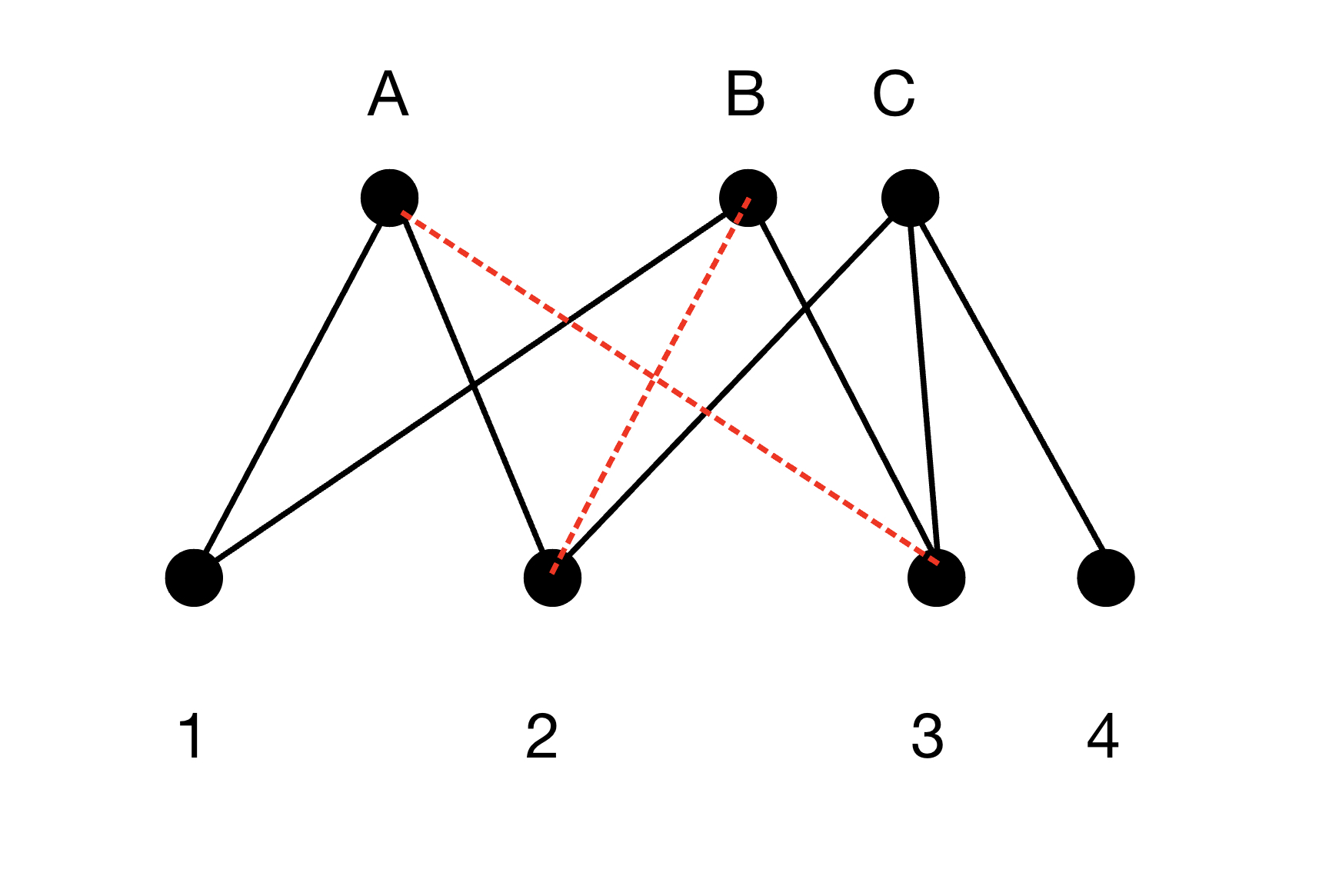}
    \caption{Example of internal links within a bipartite graph (dotted red edges).}
\end{figure}

One idea used in weighted link prediction is the Pearson coefficient \cite{pearson}. The Pearson coefficient does not predict edges, but it determines a similarity between two nodes within the same disjoint bipartite set. Similarity is determined by first finding the set of common neighbors between the two nodes. Then, for each of the two nodes, the weight of the edge with the common neighbor is subtracted with the average weight of all edges extending from that node. These values are multiplied and this process is repeated for all common neighbors. In the end, if the resulting value is positive, it will correspond to a positive similarity and vice versa.

However, it's also important to note the difference between link prediction and link recommendation. Link prediction deals with finding any arbitrary connections that will occur in the future. Link recommendation is more specific to the weighted bipartite graph. Instead of dealing with arbitrary connections, link recommendation finds the most relevant connections between unpaired top and bottom nodes and calculates a weight for a hypothetical edge. 

Specifically with regard to link recommendation, a classical approach has been collaborative filtering \cite{wiki}. The underlying principle behind collaborative filtering is very similar to that of the Pearson coefficient: if Person A and B rate a product similarly, then Person A should rate another product more similar to Person B than a randomly chosen person. Thus, collaborative filtering utilizes the power of people with mutual interests to generate recommendations. 

There are many different ways that collaborative filtering can be applied: for example, through using neural networks \cite{neural} to establish deeper analysis between the user-user and user-item connections, and hybrids between memory and matrix based ideas \cite{rec}. Collaborative filtering is not the only approach to link recommendation \cite{simpaper}, but the role of similarity still plays a vital role.

Overall, for the link recommendation problem, the error with entirely random predictions would be around 33.3\% \cite{jester}. Traditionally, important factors that affect recommendation algorithms are the size and the density of the dataset\cite{factors}. This paper builds off the previous work of both past link prediction and recommendation models to create an algorithm suited towards recommendation. 

\section{Methodology}

There are two steps to this recommendation algorithm. To make things more intuitive, we can also think of top nodes as products and bottom nodes as customers. For ease of understanding, bottom node/customer and top node/product will be used interchangeably. Thus, the weights between top and bottom node represent the rating a customer has given a product.

First, we must decide which pairs of currently nonadjacent top and bottom nodes to make a prediction. Second, after we decide upon the pair, we must actually make a prediction.

A recommendation between a top node and bottom node cannot be made if there is insufficient data connecting them. For each candidate pairing, let $t$ denote the number of neighbors of the original top node, excluding the original bottom node. Next, for each of these neighbors, let $n$ define the number of neighbors that have at least one common top neighbor with the original bottom node. If we interpret the fraction $\frac{n}{t}$, a value close to 1 means that most of the customers who already purchased the candidate product have purchased items the candidate customer has already purchased. The inverse is also true. Thus, it remains to define the threshold for the value $\frac{n}{t}$ such that all values below classify as "insufficient" for recommendation and all values above would classify as "sufficient" for recommendation.

This threshold should vary for every dataset. Generally speaking, the threshold should be lower in datasets with a sparser distribution of edges and greater in more compact datasets. Additionally, datasets with many edges may be "sparse" if they also have many nodes. 

Thus, we let the threshold be given by the following equation where $x$ represents the average amount of edges a bottom node has and $y$ represents the average amount of edges a top node has: $\frac{9}{10} - \frac{4}{x+y}$.
We set this threshold to have a maximum value of $\frac{9}{10}$. Thus, for larger values of $x+y$, the candidate bottom node must have a common neighbor with at least 90\% of candidate-top adjacent bottom nodes. The term $-\frac{4}{x+y}$ serves to compensate for smaller datasets with fewer existing edges. The constant $4$ is specifically chosen to accommodate smaller datasets such as the Epinions dataset. After performing some initial testing, the constant values $1$, $2$, and $3$ all return little or none recommendations because the resultant threshold is too high. The constant value $4$ is the first number that consistently produced recommendations.

However, this does not prevent cases where the product and customer have sufficient similarity, but have insufficient data to ascertain that level of sufficiency. For example, if candidate top node has only one bottom neighbor besides the candidate bottom node and the candidate bottom node has a common top neighbor with the one candidate top-adjacent bottom node, then $\frac{n}{t}=1$. This will always be above the threshold, but since there is only one data point, it is too few to tell. As a result, we also require that the original bottom node and bottom nodes adjacent to the original top node have a number of common neighbors at least equal to the average edges of a top node.

Once we have filtered out the unpaired nodes with insufficient data, it remains to make a prediction for the expected weight between the candidate top node and the candidate bottom node. This is achieved by returning to the bottom nodes that are neighbors of the candidate top node. For the sake of clarity, we label these bottom nodes as candidate top adjacent. We define the \emph{similarity} of one of these bottom nodes and the candidate bottom node to be how closely they rate their common neighbors. In other words, the \emph{similarity} is how closely the preferences of two customers match, based on the products they both have bought. Naturally, a higher similarity means that two customers have similar preferences. In turn, it affects the prediction positively. 

We determine similarity by comparing all products rated by both. For each common product, we compute the difference between how the candidate bottom node has rated it compared to the average of how the common product has been rated by all of its customers overall. We do the same for the other bottom node. The results of these differences are compared, and a temporary similarity for each common top node between the candidate top adjacent bottom node and the candidate bottom node is calculated. This is based on their absolute difference and whether their signs are the same. The piecewise function in Figure 3 depicts exactly how temporary similarity is calculated ($r_1$ and $r_2$ denote the two differences; $a$ denotes the average rating of the common top node). 

\begin{figure}
    \centering
    \includegraphics[width=9cm]{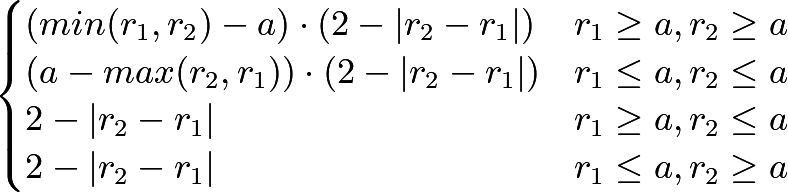}
    \caption{Piecewise function for temporary similarity}
\end{figure}

After iterating through each of the common products, we take the average of the temporary similarity values to find the overall similarity between the candidate bottom node and the candidate top-neighboring bottom node.

For each of the bottom nodes neighboring the candidate top node, we compute its similarity with the candidate bottom node using the above algorithm. Then, our predicted rating $p$ is determined by the following formula:

$k$ represents the number of top-adjacent bottom nodes that share at least one common neighbor with the candidate bottom node. $s_i$ represents the similarity score for the ith product-adjacent bottom node. $r_i$ is the analogue for rating.

$$p=\frac{\Sigma^{k}_{n=1} s_i\cdot r_i}{\Sigma^{k}_{n=1} s_i}$$
\section{Results}

We test the algorithm using two datasets: a dataset scrapped from the Epinions website \cite{epinions} and a dataset collected through the MovieLens website \cite{movielens}. The Python code can be found in the Appendix section. Both datasets can be modeled with weighted bipartite graphs with weights ranging between 1 and 5. Within the Epinions dataset, users are rating products. Within the Movielens dataset, users are rating movies. In both datasets, a 1 represents an utterly unsatisfactory response whereas a 5 represents an extremely satisfactory response.

After reading in the dataset, we use 80\% of the given edges to train the algorithm (providing the averages used in later calculations) and the remaining 20\% of edges were set aside to test the algorithm. 

The algorithm is tested by first running through 20\% of edges initially set aside and the first part of the algorithm determined whether the edge has sufficient data in order to make a prediction. If the edge has sufficient data, the second part of the algorithm is implemented to obtain a predicted weight.

\begin{figure}[H]
    \centering
    \includegraphics[width=6cm]{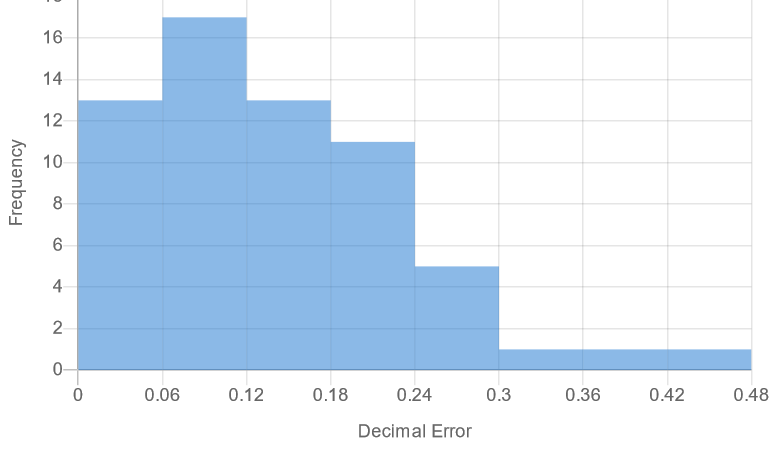}
    \caption{Histogram showing individual error of Epinions recommendations}
\end{figure}

\begin{figure}[H]
    \centering
    \includegraphics[width=6cm]{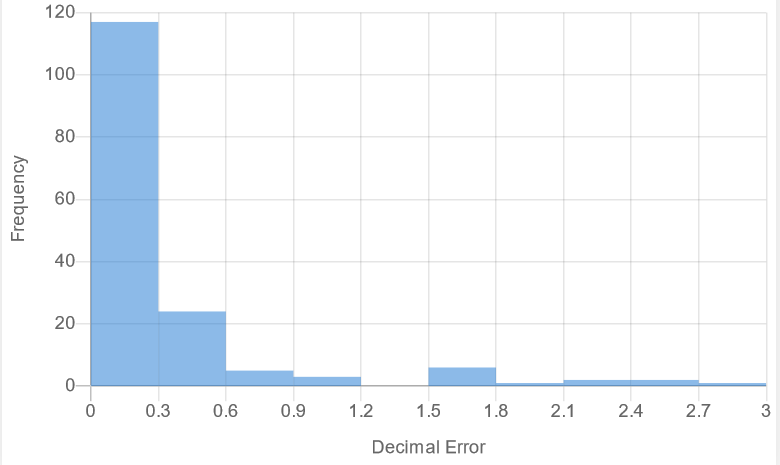}
    \caption{Histogram showing individual error of Movielens recommendations}
\end{figure}

The predicted weight is measured against the actual weight and the percent error was determined. To determine the overall percent error of our algorithm, we average all of the individual percent error values of each prediction. 

The algorithm's performance across both datasets is shown in more detail in Figures 4 and 5, and the performance is also summarized in the table below. 

\begin{center}
\begin{tabular}{|c|c|}
\hline
    \textbf{Dataset} & \textbf{Percent Error} \\
    \hline
    Epinions & 13.8\% \\
    \hline
    Movielens & 14.4\% \\
    \hline
\end{tabular}
\end{center}

Comparing Figures 4 and 5, we notice that the error in predictions from the Epinions dataset is approximately unimodal, centered between 6 and 18 percent. On the contrary, the error in predictions from the Movielens dataset is much more skewed right and has less clear of a center value. Thus, when determining the overall percent error, we took the mean for the Epinions dataset and the median from the Movielens dataset.

Overall, our percent error for the two datasets hovers at around 14\%. This is a promising number. However, it must be remembered that each of our recommendations was "screened", meaning that a recommendation was only made if a sufficient amount of data was there in the first place.

\section{Discussion}

Our algorithm works fairly well, but there exist noticeably many outlier recommendations, noticeably more from the Movielens dataset. Specifically, there are error values up to 300\% generated from the Movielens recommendations. Analyzing this, since both datasets employ a similar 1-5 rating system and also model some kind of product-customer relation, the largest difference between the two datasets is that the Movielens dataset has almost 10 times the amount of entries as the Epinions dataset. The difference in size perhaps leads to a larger diversity of opinions which in turn demands a different approach to our parameters.

Related to the size of datasets, this algorithm runs in $O(n^3)$ time. Therefore, it becomes infeasible to test the algorithm on datasets with anything larger than approximately $100,000$ entries without running into an unrealistic run time. A significant amount of existing datasets have entries in the millions, so there are limitations in the scope of datasets we are able to test upon. However, for their size, the Epinions and Movielens datasets both are a solid representation of a real world product-customer relationship.

Looking back, the way the $\frac{n}{t}$ value is calculated can also be put into question. In our algorithm, we define $n$ as the number of candidate bottom node and top-adjacent bottom nodes that share at least one top node in common. But, instead of one common top node, we may want the two bottom nodes to share at least two common top nodes, three common top nodes, or even more common top nodes.

However, it also must be noted that if this parameter were to be adjusted, the $\frac{n}{t}$ threshold would likely also have to change. For requirements of more common top nodes, the value of $n$ will decrease and the $\frac{n}{t}$ threshold would have to lower correspondingly. This makes sense: while there might be less pairs of sufficient bottom nodes, we would ultimately have relatively equal data to work with because of the requirements for more common top nodes.

Furthermore, we can also tinker with the conditions in which the $\frac{n}{t}$ value is considered sufficient enough to generate a prediction. Looking at the threshold, potential edges are sometimes rejected because data is insufficient. Although the $\frac{n}{t}$ value is high enough, the value $t$ might have been too low for us to be confident in making a prediction. However, instead of outright rejecting those edges, we could adjust the threshold so that it would increase as the number of top-adjacent nodes decrease.

Finally, we can also explore different functions that determine similarity between the candidate bottom node and a top-adjacent bottom node. In a manner similar to how the Pearson coefficient was calculated, the piecewise function in Figure 3 rewards the bottom nodes for rating their common top nodes similarly. It does this by taking into account the differences between how the two bottom nodes rated their common top nodes and how they typically rated a top node. If the signs match (positive or negative), then it would contribute more positively towards similarity. However, we could explore the effect of slightly modifying the constants or introducing more complexity into the functions. 

\section{Conclusion}

In this paper, we present another approach to the classical link recommendation problem. Extending ideas from previous papers that focused on sign prediction \cite{pearson} or link prediction \cite{LinkReference1}, we utilize the properties of the weighted bipartite graph to determine two things: the most relevant edges that could appear based on edges that already existed as well as what the weights of those edges would be based on the idea of \emph{similarity}. 

For the first step, we consider whether the candidate bottom node had common neighbors with bottom nodes adjacent to the candidate top node. We compare this number against the total number of bottom nodes adjacent to the candidate top node. After checking this, we check if any edges considered sufficient are a result of low sample size. The exact tuning parameters are determined by the density and size of the dataset.

For the second step, we use the idea of \emph{similarity}. Similarity is calculated between the candidate bottom node and candidate top adjacent bottom nodes by taking into account how closely they rated common neighbors. In the end, the weight of the edge between the candidate bottom node and the candidate top is more heavily influenced by higher similarity pairings and less influenced by lower similarity pairings.

We test our algorithm upon the Movielens and Epinions datasets, and find that our algorithm performs quite well compared with existing algorithms \cite{jester}. Nevertheless, we can keep working to refine and improve the results of the proposed algorithm in this paper.

\section{Acknowledgements}

I would like to thank Professor Zhiliang Xu, at the Department of Applied and Computational Mathematics and Statistics at the University of Notre Dame, for introducing this topic to me, and for providing invaluable feedback during the project and during the writing of this paper. 

\printbibliography

@book{Read_Atlas,
author = {Read, Ronald C. and Wilson, Robin J.},
title = {An Atlas of Graphs (Mathematics)},
year = {2005},
publisher = {Clarendon Press}
}

@article{movielens,
  author = {Harper, F Maxwell and Konstan, Joseph A},
  journal = {ACM Transactions on Interactive Intelligent Systems},
  number = 4,
  pages = {1--19},
  publisher = {ACM New York, NY, USA},
  title = {The MovieLens Datasets: History and Context},
  volume = 5,
  year = 2015
}

@INPROCEEDINGS{LinkReference1,
  author={Allali, Oussama and Magnien, Clémence and Latapy, Matthieu},
  booktitle={2011 IEEE Conference on Computer Communications Workshops (INFOCOM WKSHPS)}, 
  title={Link prediction in bipartite graphs using internal links and weighted projection}, 
  year={2011},
  volume={},
  number={},
  pages={936-941},
  doi={10.1109/INFCOMW.2011.5928947}}

@MISC{epinions,
title = {Dataset collected by Paolo Massa},
howpublished={\url{http://www.trustlet.org/downloaded_epinions.html}}
}

@InProceedings{networkx,
  author =       {Aric A. Hagberg and Daniel A. Schult and Pieter J. Swart},
  title =        {Exploring Network Structure, Dynamics, and Function using NetworkX},
  booktitle =   {Proceedings of the 7th Python in Science Conference},
  pages =     {11 - 15},
  address = {Pasadena, CA USA},
  year =      {2008},
  editor =    {Ga\"el Varoquaux and Travis Vaught and Jarrod Millman},
}

@book{social,
title = "Social Network Data Analytics",
editor = "Charu C. Aggarwal",
year = "2011",
publisher = "Springer",
ADDRESS = {New York, NY},
}

@book{graph,
title = "Graph Theory With Applications",
author = "J. A. Bondy and U. S. R. Murty",
year = "1976",
publisher = "Macmillan Press",
ADDRESS = {Great Britain},
}

@article{jester,
  title={Eigentaste: A Constant Time Collaborative Filtering Algorithm},
  author={Ken Goldberg and Theresa Roeder and Dhruv Gupta and Chris Perkins},
  journal={Information Retrieval},
  year={2004},
  volume={4},
  pages={133-151}
}

@MISC{wiki,
title = {Collaborative filtering},
howpublished={\url{https://en.wikipedia.org/wiki/Collaborative_filtering}}
}

@book{pearson,
  author={Rahman, N. A.},
  title={A Course in Theoretical Statistics},
  year=1968,
  publisher= "Charles Griffin and Company",
}

@inproceedings{neural,
title = "Neural Collaborative Filtering ",
year = "2017",
author = " Xiangnan He and Lizi Liao and Hanwang Zhang and Liqiang Nie and Xia Hu and Tat-Seng Chua",
month = "April",
doi = "http://dx.doi.org/10.1145/3038912.3052569",
booktitle = "WWW '17: Proceedings of the 26th International Conference on World Wide Web"
}

@INPROCEEDINGS{rec,
  author={Zhang, Ruisheng and Liu, Qi-dong and Chun-Gui and Wei, Jia-Xuan and Huiyi-Ma},
  booktitle={2014 Second International Conference on Advanced Cloud and Big Data}, 
  title={Collaborative Filtering for Recommender Systems},
  year={2014},
  volume={},
  number={},
  pages={301-308},
  doi={10.1109/CBD.2014.47}}

@INPROCEEDINGS{simpaper,
  author={Ghaleb, Hamid and Abdullah-Al-Wadud, M.},
  booktitle={2019 International Conference on Sustainable Technologies for Industry 4.0 (STI)}, 
  title={An Enhanced Similarity Measure for Collaborative Filtering-based Recommender Systems}, 
  year={2019},
  volume={},
  number={},
  pages={1-4},
  doi={10.1109/STI47673.2019.9068084}}

@PHDTHESIS{factors,
  title    = "Evaluating Prediction Accuracy for Collaborative Filtering
              Algorithms in Recommender Systems",
  author   = "Salam Patrous, Ziad and Najafi, Safir",
  year     =  2016,
  url      = "http://kth.diva-portal.org/smash/record.jsf?aq2=%5B%5B%5D%5D&c=1&af=%5B%5D&searchType=LIST_LATEST&query=&language=en&pid=diva2%3A927356&aq=%5B%5B%5D%5D&sf=all&aqe=%5B%5D&sortOrder=author_sort_asc&onlyFullText=false&noOfRows=50&dswid=-7195",
  address  = "Stockholm, Sweden",
  school   = "KTH Royal Institute of Technology"
}

\newpage
\section{Appendix}
\noindent
    \sentence
\begin{lstlisting}[language=Python, basicstyle=\ttfamily\scriptsize]
import networkx as nx
import random
from networkx import common_neighbors
from networkx.algorithms import bipartite
f = open("movielens.txt", "r")
tr_G = nx.Graph()
te_G = nx.Graph()
tr_top_nodes = set()
tr_bottom_nodes = set()
te_top_nodes = set()
te_bottom_nodes = set()
ct = 0
for x in f:
    r_num = random.randint(1, 5)
    nums = x.split()
    product = "-" + nums[1]
    if r_num == 1:
        te_G.add_node(nums[0], bipartite=0)
        te_G.add_node(product, bipartite=1)
        te_G.add_edge(nums[0], product, weight=int(nums[2]))
        te_bottom_nodes.add(nums[0])
        te_top_nodes.add(product)
    else:
        tr_G.add_node(nums[0], bipartite=0)
        tr_G.add_node(product, bipartite=1)
        tr_G.add_edge(nums[0], product, weight=int(nums[2]))
        tr_bottom_nodes.add(nums[0])
        tr_top_nodes.add(product)
top_avg = {}
for i in tr_top_nodes:
    sum = 0
    ct = 0
    for j in tr_G.neighbors(i):
        sum += tr_G.get_edge_data(i, j).get("weight")
        ct += 1
    top_avg[i] = sum/ct

error = 0
actual = 0
test = 0
count = 0
for x, y, a in te_G.edges(data=True):
    num = 0
    tot = 0
    if int(x)>0:
        u = x
        v = y
    else:
        u = y
        v = x
    if v in tr_G and u in tr_G:
        for k in tr_G.neighbors(v):
            if k != u:
                ct = 0
                tot += 1
                for i in common_neighbors(tr_G, u, k):
                    if i != v:
                        ct += 1
                if ct > 0:
                    num += 1
        if tot > 59 and num / tot >= 9/10:
            sum = 0
            num = 0
            for k in tr_G.neighbors(v):
                if k!=u:
                    neighbors = 0
                    sim = 0
                    for i in common_neighbors(tr_G, u, k):
                        neighbors += 1
                        avg = top_avg[i]
                        r_1 = tr_G.get_edge_data(u, i).get("weight")
                        r_2 = tr_G.get_edge_data(k, i).get("weight")
                        if r_1 > avg and r_2 > avg:
                            sim += (min(r_2, r_1) - avg) * (2 - abs(r_2 - r_1))
                        elif r_1 < avg and r_2 < avg:
                            sim += (avg - max(r_2, r_1)) * (2 - abs(r_2 - r_1))
                        else:
                            sim += 2 - abs(r_2 - r_1)
                    if neighbors > 0:
                        sim = sim / neighbors
                        num += sim
                        sum += (tr_G.get_edge_data(v, k).get("weight") * sim)
            rating = sum / num
            int diff = abs(rating-int(te_G.get_edge_data(u, v).get("weight")))
            int actual = int(te_G.get_edge_data(u, v).get("weight"))
            test += diff/actual
            count += 1
print(test/count)
\end{lstlisting}

\end{document}